\documentclass{article}

\usepackage{color}
\usepackage{amssymb}

\newcommand{\be}{\begin{equation}}
\newcommand{\ee}{  \end{equation}}
\newcommand{\ba}{\begin{eqnarray}}
\newcommand{\ea}{  \end{eqnarray}}

\newcommand{\bas}{\begin{eqnarray*}}
\newcommand{\eas}{  \end{eqnarray*}}

\begin{document}

\title{Massive Modes for Quantum Graphs}

\author{H. A. Weidenm{\"u}ller \\
  Max-Planck-Institut f{\"u}r Kernphysik, 69029 Heidelberg, Germany}

\maketitle

\begin{abstract}
  The spectral two-point function of chaotic quantum graphs is
  expected to be universal. Within the supersymmetry approach, a proof
  of that assertion amounts to showing that the contribution of
  non-universal (or massive) modes vanishes in the limit of infinite
  graph size. Here we pay particular attention to the fact that the
  massive modes are defined in a coset space. Using the assumption
  that the spectral gap of the Perron-Frobenius operator remains
  finite in the limit, we then argue that the massive modes are indeed
  negligible.
\end{abstract}

\section{Motivation}
\label{mot}

According to the Bohigas-Giannoni-Schmit (BGS)
conjecture~\cite{Boh84}, the spectral fluctuation properties of
Hamiltonian systems that are chaotic in the classical limit, coincide
with those of the random-matrix ensemble in the same symmetry class
(unitary, orthogonal, symplectic). Numerical simulations (see, f.i.,
Ref.~\cite{Kot99}) strongly suggest that the BGS conjecture holds
likewise for chaotic quantum graphs. Analytical arguments in support
of the BGS conjecture for chaotic quantum graphs have been presented
in several papers~\cite{Gnu04, Gnu05, Plu14, Plu15}. All these
approaches use the supersymmetry formalism and a separation of the
modes of the system into the universal (or massless or zero) mode and
a number of massive modes. An essential part of the argument then
consists in showing that the contribution of massive modes to all
correlation functions vanishes in the limit of infinite graph size
(number $B$ of bonds to $\infty$). That leaves only the contribution
of the zero mode, and universality of all correlation functions
follows. The zero mode and the massive modes range in a non-linear
space of cosets. That fact has not been addressed explicitly in
previous work~\cite{Gnu04, Gnu05, Plu14, Plu15}. The problem has not
gone unnoticed. Indeed, in Ref.~\cite{Alt15} mathematical aspects of
the non-linearity of the coset space were discussed in detail without,
however, establishing control of the contribution from the massive
modes.

In the present paper we aim at filling that gap. We introduce the zero
mode and the massive modes in a manner that is consistent with the
non-linear coset structure. We do so using the strong assumption that
in the limit $B \to \infty$, the spectrum of the Perron-Frobenius
operator (defined in Eq.~(\ref{2}) below) has a finite gap that
separates the eigenvalue $\lambda_1 = 1$ from the rest. In contrast,
Refs.~\cite{Gnu04} and \cite{Gnu05} pose only the weaker condition
that the gap closes (for $B \to \infty$) no faster than $B^{- \alpha}$
with $0 \leq \alpha < 1/2$. That is perhaps a realistic requirement:
Numerical simulations~\cite{Kot99} suggest that chaotic graphs obeying
Kirchhoff boundary conditions at each vertex possess universal
spectral correlations even though the gap closes for large $B$.
However, we doubt that, in the framework of a perturbative treatment
of massive modes, there exists an easy way to turn the reasoning of
Refs.~\cite{Gnu04, Gnu05} into a convincing argument showing that the
contribution of the massive modes vanishes in all orders. This is why
we settle for the stronger assumption of a gap that remains finite in
the limit of infinite graph size $B$. That assumption prevents us from
keeping the local structure of the quantum graph fixed in the limit $B
\to \infty$: To prevent the gap from closing, the graph connectivity
must increase as the graph size is taken to infinity.

In Section~\ref{two} we recall some basic facts on chaotic
graphs~\cite{Kot99}. Section~\ref{eff} forms the central piece of the
paper. We introduce the universal mode and the massive modes in a
manner that is consistent with the coset structure, and we express the
effective action in terms of these variables. We use the fact that the
spectrum of the Perron-Frobenius operator possesses a gap. In
Section~\ref{eva} we address the superintegrals over massive modes,
and we argue that the contribution of these modes to the two-point
correlation function vanishes in the limit of infinite graph size. We
do so under rather restrictive assumptions concerning the matrix
${\cal B}$ that describes amplitude propagation on the quantum graph,
see Section~\ref{two}. We assume that the elements of ${\cal B}$ are
all of order $1 / \sqrt{2 B}$, see Eq.~(\ref{59}). That implies that
the fluctuations of the matrix elements about their mean values
(defined by the unitarity of ${\cal B}$) are small (of order $1 /
\sqrt{2 B}$). For a general proof of the BGS conjecture it would be
necessary to lift that assumption.

We confine ourselves to the unitary case, to closed graphs, and to the
two-point function (the correlation function of the retarded and the
advanced Green's function).

\section{Two-Point Function}
\label{two}

To set the stage, we briefly summarize previous work~\cite{Kot99,
  Gnu04, Gnu05, Plu14, Plu15} on chaotic quantum graphs. We consider
connected simple graphs with $V$ vertices and $B$ bonds. Each bond has
two directions $d = \pm$. The $2 B$ directed bonds are labeled $(b
d)$. On every directed bond the Schr{\"o}dinger wave carries the same
wave number $k$ and a direction-dependent magnetic phase $\phi_{b d}$
that breaks time-reversal invariance. Hermitean boundary conditions at
vertex $\alpha$ (with $\alpha = 1, 2, \ldots, V$) cause incoming and
outgoing waves on the bonds linked to $\alpha$ to be related by a
unitary vertex scattering matrix $\sigma^{(\alpha)}$. When arranged in
directed-bond representation, the $V$ scattering matrices
$\sigma^{(\alpha)}$ form the $(2 B)$-dimensional unitary bond
scattering matrix $\Sigma^{(B)}$ with elements $\Sigma^{(B)}_{b d, b'
  d'}$. Amplitude propagation within the graph depends upon the
unitary matrix
\be
{\cal B}_{b d, b' d'} = \Big( \sigma^D_1 \Sigma^{(B)} \Big)_{b d, b' d'} \ .
\label{1}
\ee
Here $\sigma^D_1$ is the first Pauli spin matrix in two-dimensional
directional space. It flips the direction of bonds, $( \sigma^D_1
\Sigma^{(B)} )_{b d, b' d'} = \Sigma^{(B)}_{b - d, b' d'}$. To see how
$\sigma^D_1$ arises (see Refs.~\cite{Plu14, Plu15}) we consider a bond
connecting vertices $\alpha$ and $\beta$. For vertex $\alpha$ the bond
is denoted by $(\alpha, \beta) = (b d)$, for vertex $\beta$ the bond
is denoted by $(\beta, \alpha) = (b, -d)$. The two bond directions
differ. To correctly describe amplitude propagation through the graph,
the bond directions must match. That is achieved by multiplying
$\Sigma^{(B)}$ with $\sigma^D_1$. The graph is classically chaotic if
in the limit $B \to \infty$ of infinite graph size the spectrum of the
Perron-Frobenius operator, i.e., of the matrix
\be
{\cal F}_{b d, b'd'} = | {\cal B}_{b d, b'd'} |^2
\label{2}
\ee
possesses a finite gap separating the leading eigenvalue $\lambda_1 =
1$ from all other eigenvalues $\lambda_i$ (so that $|\lambda_i| \leq
(1 - a)$ with $a > 0$). That is assumed throughout.

Unitary symmetry is realized by averaging separately and independently
over the phases $\phi_{b d}$ ranging in the interval $[0, 2 \pi]$. The
averages are carried out using the supersymmetry method and the
color-flavor transformation~\cite{Zir96}. As a result, the two-point
function is written as the derivative of a generating function, an
integral in superspace. That function carries in the exponent the
effective action ${\cal A}_{\rm eff}$ given here in the form of
Ref.~\cite{Plu15} (see also Refs.~\cite{Gnu04, Gnu05}),
\be
{\cal A}_{\rm eff} = - {\rm STr} \ln (1 - Z \tilde{Z}) + {\rm STr}
\ln (1 - w_+ {\cal B}_+ Z {\cal B}^\dag_- w_- \tilde{Z} ) \, .
\label{3}
\ee
Here $w_+ = w^*_- = \exp \{ i \kappa {\cal L} \}$ where ${\cal L}$ is
the diagonal matrix of bond lengths $L_b$, and where $\kappa$ ($-
\kappa$) is the wave number increment in the retarded sector (the
advanced sector, respectively). The matrices ${\cal B}_\pm$ are
defined as
\be
{\cal B}_\pm = (1 +  j_\pm \frac{i \pi}{B} \sigma^s_3) {\cal B} \, ,
\label{4}
\ee
with $\sigma^s_3$ the third Pauli spin matrix in superspace. The
average two-point function is obtained by differentiation of the
generating function with respect to $j_+$ and $j_-$ at $j_+ = 0 =
j_-$.

The supermatrices $Z = \{ \delta_{b b'} \delta_{d d'} Z_{b d; s s'}
\}$ and $\tilde{Z} = \{ \delta_{b b'} \delta_{d d'} \tilde{Z}_{b d; s
  s'} \}$ are diagonal in directed-bond space. For fixed indices $(b
d)$ the matrices $Z$ and $\tilde{Z}$ each have dimension two and form
part of a supermatrix of dimension four,
\be
\left( \matrix{ 0 & Z \cr
               \tilde{Z} & 0 \cr} \right) .
\label{5}
\ee
In Boson-Fermion block notation we have
\be
Z = \left( \matrix{ Z_{B B} & Z_{B F} \cr
               Z_{F B} & Z_{F F} \cr} \right) , \quad
\tilde{Z} = \left( \matrix{ \tilde{Z}_{B B} & \tilde{Z}_{B F} \cr
               \tilde{Z}_{F B} & \tilde{Z}_{F F} \cr} \right) ,
\label{6}
\ee
where
\be
Z^{}_{B B} = \tilde{Z}^*_{B B} \, , \ Z^{}_{F F} = - \tilde{Z}^*_{F F} \,
, \ | Z^{}_{B B} | < 1 \, .
\label{7}
\ee
The variable transformation $(1 + j_+ \frac{i \pi}{B} \sigma^s_3) Z
\to Z$, $(1 + j_- \frac{i \pi}{B} \sigma^s_3) \tilde{Z} \to \tilde{Z}$
with Berezinian unity is used to simplify the source terms and, after
differentiation with respect to $j_+$ and $j_-$, yields for the
effective action
\be
{\cal A} = - {\rm STr} \ln (1 - Z \tilde{Z}) + {\rm STr}
\ln (1 - w_+ {\cal B} Z {\cal B}^\dag w_- \tilde{Z} )
\label{11}
\ee
and for the source terms
\ba
&& \frac{\pi^2}{B^2} \Big( {\rm STr} \big[ \sigma^s_3
(1 - Z \tilde{Z})^{- 1} Z \tilde{Z} \big] \, {\rm STr} \big[ \sigma^s_3
(1 - \tilde{Z} Z)^{- 1} \tilde{Z} Z \big] \nonumber \\
&& \hspace{1.3cm} + {\rm STr} \big[ \sigma^s_3 Z (1 - \tilde{Z} Z)^{- 1}
\sigma^s_3 \tilde{Z} (1 - Z \tilde{Z})^{- 1} \big] \Big) \, .
\label{12}
\ea
The terms~(\ref{12}) multiply $\exp \{ - {\cal A} \}$ (Eq.~(\ref{11}))
under the superintegral over $(Z, \tilde{Z})$ with flat measure. That
superintegral constitutes an exact representation of the average
two-point function.

\section{Effective Action}
\label{eff}

We introduce the universal mode and the massive modes. We express the
effective action~(\ref{11}) and the source terms ~(\ref{12}) as
functions of these modes.

\subsection{Coset Space}
\label{coset}

We first focus attention on the bare effective action, obtained from
Eq.~(\ref{11}) by omission of $\kappa$, i.e., by putting $w_+ = 1 =
w_-$,
\be
{\cal A}_{\rm bare}(\tilde{Z}, Z) = - {\rm STr} \ln (1 -  Z \tilde{Z})
+ {\rm STr} \ln (1 - {\cal B} Z {\cal B}^\dag \tilde{Z} ) .
\label{13}
\ee
Using an argument of Ref.~\cite{Alt15} we show that ${\cal A}_{\rm
  bare}(\tilde{Z}, Z)$ is defined in a coset space. In
retarded-advanced notation we define the matrices
\be
\Lambda = \left( \matrix{ 1 & 0 \cr
                          0 & - 1 \cr} \right) , \quad
M = \left( \matrix{ {\cal B} & 0 \cr
                    0 & {\cal B}^\dag \cr} \right) .
\label{14}
\ee
We expand the logarithms in Eq.~(\ref{13}), resum and obtain
\be
{\cal A}_{\rm bare} = - {\rm STr} \ln \frac{2}{1 + Q \Lambda} +
{\rm STr} \ln \bigg( 1 - M \frac{Q \Lambda - 1}{Q \Lambda + 1}
\bigg)
\label{15}
\ee
where
\ba
Q &=& \left( \matrix{
(1 + Z \tilde{Z})(1 - Z \tilde{Z})^{- 1} & - 2 Z (1 - \tilde{Z}
Z)^{- 1} \cr 2 \tilde{Z}(1 - Z \tilde{Z})^{- 1} & - (1 + \tilde{Z}
Z)(1 - \tilde{Z} Z)^{- 1} \cr} \right) \nonumber \\
&=& g(Z) \Lambda (g(Z))^{- 1}
\label{16}
\ea
and
\be
g(Z) = \left( \matrix{
     (1 - Z \tilde{Z})^{- 1/2} & Z (1 - \tilde{Z} Z)^{- 1/2} \cr
  \tilde{Z} (1 - Z \tilde{Z})^{- 1/2} & (1 - \tilde{Z} Z)^{- 1/2}
                        \cr} \right) .
\label{17}
\ee
The supermatrices $Z$ and $\tilde{Z}$ are diagonal in directed bond
space and so is, therefore, $Q$.

We consider a specific set of bond indices $(b d)$, omit these and
work in four-dimensional superspace only. The matrix $Q$ and the bare
action ${\cal A}_{\rm bare}$ remain unchanged under the transformation
$g \to g k$ if $k$ commutes with $\Lambda$. Therefore, $Q$ and ${\cal
  A}_{\rm bare}$ are defined in the coset superspace $G / K$ where $g
\in G = U(1, 1|2)$ and $k \in K = U(1|1) \times U(1|1)$ with
fundamental form $Q = g \Lambda g^{- 1} = g k \Lambda k^{- 1} g^{-
  1}$. Writing
\be
g = \left( \matrix{ A & B \cr C & D \cr} \right)
\label{18}
\ee
we find from Eq.~(\ref{17})
\ba
&&Z(Q) = B(g) (D(g))^{- 1} = B(g k) (D(g k))^{- 1} , \nonumber \\
&&\tilde{Z}(Q) = C(g) (A(g))^{- 1} = C(gk) (A(gk))^{- 1}
\label{19}
\ea
for the local coordinates $(Z, \tilde{Z})$ of $Q$ used in
Eqs.~(\ref{16}). As briefly explained in the paragraph following
Eq.~(\ref{37}), these and the other local coordinates introduced below
are, however, not globally defined.

A group element $g_0 \in G$ acts on $Q$ by $Q \to g_0 Q g^{- 1}_0$
and on $g \in G $ by left multiplication. With
\be
g_0 = \left( \matrix{ A_0 & B_0 \cr C_0 & D_0 \cr} \right) \ {\rm and} \
g_0 g = \left( \matrix{ A_0 A + B_0 C & A_0 B + B_0 D \cr
                       D_0 C + C_0 A & D_0 D + C_0 B \cr} \right)
\label{20}
\ee
we obtain from Eqs.~(\ref{19}) for the local coordinates of $g_0 Q
g^{- 1}_0$
\ba
&&g_0 \cdot Z(Q) = (A_0 Z(Q) + B_0)(D_0 + C_0 Z(Q))^{- 1} ,
\nonumber \\
&&g_0 \cdot \tilde{Z}(Q) = (D_0 \tilde{Z}(Q) + C_0)(A_0 + B_0
\tilde{Z}(Q))^{-1} .
\label{21}
\ea

\subsection{Zero Mode plus Fluctuations}
\label{zero}

Let $g_0 \in G$ in Eq.~(\ref{20}) be independent of the directed-bond
indices $(b d)$, and $Q_0 = g_0 \Lambda g^{- 1}_0$. Following
Eqs.~(\ref{21}) we define the zero mode (or universal mode) $(Y,
\tilde{Y})$ by the local coordinates of $Q_0$,
\be
Y = g_0 \cdot 0 = B_0 D^{- 1}_0 \ , \ \tilde{Y} = g_0 \cdot \tilde{0} =
C_0 A^{- 1}_0 \ .
\label{22}
\ee
Fluctuations about the zero mode are written in the sense of
Eq.~(\ref{21}) as
\be
Z_{b d} = g_0 \cdot \zeta_{b d} \, , \quad \tilde{Z}_{b d} =
g_0 \cdot \tilde{\zeta}_{b d} \, .
\label{28}
\ee
Eqs.~(\ref{28}) transform the local variables $(Z, \tilde{Z})$ to the
local variables $(Y, \tilde{Y})$ and $(\zeta, \tilde{\zeta})$. The
latter are not gauge invariant. Therefore, we replace the
supermatrices $(\zeta, \tilde{\zeta})$ by the gauge-invariant
supermatrices $(\xi, \tilde{\xi})$ defined by
\be
\xi = A_0 \zeta D^{- 1}_0 \, , \quad
\tilde{\xi} = D_0 \tilde{\zeta} A^{- 1}_0 \ .
\label{35}
\ee
Suppressing the directed-bond indices we use Eqs.~(\ref{21}) (with
$\zeta$ for $Z(Q)$ and $\tilde{\zeta}$ for $\tilde{Z}(Q)$) to express
$Z$ and $\tilde{Z}$ as given by Eqs.~(\ref{28}) as
\ba
&Z = g_0 \cdot \zeta = (A_0 \zeta + B_0)(D_0 + C_0 \zeta)^{- 1}
= (Y + \xi) (1 + \tilde{Y} \xi)^{- 1} , \nonumber \\
&\tilde{Z} = g_0 \cdot \tilde{\zeta} =
(D_0 \tilde{\zeta} + C_0) (A_0 + B_0 \tilde{\zeta})^{- 1} =
(\tilde{Y} + \tilde{\xi}) ( 1 + Y \tilde{\xi})^{- 1} .
\label{37}
\ea
The transformation~(\ref{37}) from the variables $(Z, \tilde{Z})$ to
the variables $(Y, \tilde{Y})$ and $(\xi, \tilde{\xi})$ is somewhat
analogous to the transformation from independent coordinates to
center-of-mass and relative coordinates. Actually, the
transformation~(\ref{37}) is more complicated than that because it has
to respect the coset structure. We very briefly remark on the ensuing
difficulties. The variables for our coset space $G/K$ cannot be
introduced in a globally well-defined way without using the
mathematical machinery of an atlas of coordinate charts, transition
functions, etc. Therefore, the gauge-independent variables $(Z,
\tilde{Z})$, $(Y, \tilde{Y})$ and $(\xi, \tilde{\xi})$ are not
globally defined but serve as good local coordinates. A simplification
arises because we confine ourselves to small fluctuations of $(Z,
\tilde{Z})$ about the ``center-of-mass'' coordinates $(Y,
\tilde{Y})$. That is done by linearizing the transformation~(\ref{28})
in the new variables $(\zeta, \tilde{\zeta})$. Even in the linear
regime we have to respect the coset structure, however. We are greatly
helped by our assumption that the spectrum of the Perron-Frobenius
operator has a sufficiently large gap. By that assumption, the
fluctuations of $(Z, \tilde{Z})$ about the ``center-of-mass'' coset
$g_0 K$ with coordinates $(g_0 \cdot 0, g_0 \cdot \tilde{0})$ are
small in a quantifiable sense, and the smallness allows us to treat
the relative variables $(\zeta, \tilde{\zeta})$ approximately as
vectors in the tangent space of $G/K$ at $g_0 K$. Then the variables
$(\xi, \tilde{\xi})$ also lie in a vector space. That is used in what
follows.

Constraints are needed because only $(2 B - 1)$ of the variables
$\zeta$ and of the variables $\tilde{\zeta}$ appearing in
Eqs.~(\ref{28}) (or of the variables $\xi$ and $\tilde{\xi}$ appearing
in Eqs.~(\ref{35})) are independent. Using the assumption that $(\xi,
\tilde{\xi})$ lie in a vector space we impose the constraints
\be
\sum_{b d} \xi_{b d} = 0 = \sum_{b d} \tilde{\xi}_{b d} \, .
\label{36}
\ee
The transformation~(\ref{37}) with the constraints~(\ref{36})
introduces as new integration variables the coordinates $(Y,
\tilde{Y})$ of the zero mode and the independent ones among the
variables $(\xi, \tilde{\xi})$. The Berezinian of the transformation
is unity.

\subsection{Implementation of the New Variables}

In order to express the effective action of Eq.~(\ref{11}) and the
source terms~(\ref{12}) in terms of the modes $(Y, \tilde{Y})$ and
$(\xi, \tilde{\xi})$, we derive an invariance property of the bare
effective action, adapting to the present case an argument developed
in Ref.~\cite{Zir97} for a network model of the Integer Quantum Hall
Effect. With the help of the definitions and results of
Section~\ref{coset} we show by explicit calculation that
\ba
&& {\rm STr} \ln (1 - (g_0 \cdot Z) (g_0 \cdot \tilde{Z})) - {\rm
  STr} \ln (1 - {\cal B} (g_0 \cdot Z) {\cal B}^\dag (g_0 \cdot
\tilde{Z})) \nonumber \\
&& \ \ = {\rm STr} \ln (1 - Z \tilde{Z}) - {\rm STr} \ln (1 -
      {\cal B} Z {\cal B}^\dag \tilde{Z}) \ .
\label{27}
\ea
The invariance property~(\ref{27}) holds provided that $g_0 \cdot
({\cal B} Z {\cal B}^\dag) = {\cal B} (g_0 \cdot Z) {\cal B}^\dag$.
We have not used any specific properties of the matrices $(Z,
\tilde{Z})$. Therefore, Eq.~(\ref{27}) holds also for the matrices
$(\zeta, \tilde{\zeta})$.  We use Eq.~(\ref{27}) for the bare
effective action of Eq.~(\ref{13}). We replace the variables $(Z,
\tilde{Z})$ by the transformation~(\ref{28}) and apply the invariance
property~(\ref{27}) to the resulting expression. That gives
\be {\cal A}_{\rm bare}(g_0 \cdot \zeta, g_0 \cdot \tilde{\zeta}) =
    {\cal A}_{\rm bare}(\zeta, \tilde{\zeta}) \ .
\label{23}
\ee
The matrices $A_0$ and $D_0$ in Eqs.~(\ref{35}) commute with ${\cal
  B}$.  Therefore, ${\cal A}_{\rm bare}(\zeta, \tilde{\zeta}) = {\cal
  A}_{\rm bare}(\xi, \tilde{\xi})$, and we obtain
\be
   {\cal A}_{\rm bare}(\xi, \tilde{\xi}) = - {\rm STr} \ln (1 - \xi
\tilde{\xi}) + {\rm STr} \ln (1 - {\cal B} \xi {\cal B}^\dag
\tilde{\xi}) \ .
\label{24}
\ee

We return to the full effective action~(\ref{11}). We are interested
in values of $\kappa$ that are of the order of the mean level spacing
$\Delta = \pi / \sum_b L_b$. Therefore, we expand ${\cal A}$ up to
terms of first order in $\kappa$. In the first-order terms we neglect
the fluctuations by putting all $\xi = 0 = \tilde{\xi}$. Since ${\cal
  B}$ commutes with the zero-mode variables $Y$ and $\tilde{Y}$ that
gives
\be
{\cal A} \approx {\cal A}_{\rm bare}(\xi, \tilde{\xi}) - \frac{2 i
  \pi \kappa} {\Delta} \, {\rm STr}_s \frac{1}{1 -Y \tilde{Y}} \, ,
\label{38}
\ee
with ${\cal A}_{\rm bare}$ given by Eq.~(\ref{24}). The index $s$
indicates that the supertrace extends only over superspace. Being
proportional to ${\rm STr} [g(Y) \Lambda (g(Y))^{- 1} \Lambda] = {\rm
  STr} [Q \Lambda]$, the last term in Eq.~(\ref{38}) has the classical
form for the symmetry-breaking term.

We turn to the source terms~(\ref{12}). The supermatrices $(Z,
\tilde{Z})$ are diagonal in directed bond space; we consider a fixed
value of $(b d)$ and omit these indices. Then both $Z$ and $\tilde{Z}$
have dimension two. We define the four-dimensional matrix $Q(\xi,
\tilde{\xi})$ as in Eq.~(\ref{16}) but with the replacements $Z \to
\xi$, $\tilde{Z} \to \tilde{\xi}$.  With $\Lambda$ defined in
Eqs.~(\ref{14}), the source terms are written as
\ba
&& {\rm STr}_s \big[ \sigma^s_3 (1 - Z \tilde{Z})^{- 1} Z \tilde{Z} \big] =
- 1 + {\rm STr}_s \big[ \Sigma(Y, \tilde{Y}) Q(\xi, \tilde{\xi})
  \Lambda \big] \, , \nonumber \\
&& {\rm STr}_s \big[ \sigma^s_3 (1 - \tilde{Z} Z)^{- 1} \tilde{Z} Z \big] =
- 1 + {\rm STr}_s \big[ \Sigma'(Y, \tilde{Y}) Q(\xi, \tilde{\xi})
  \Lambda \big] \, , \nonumber \\
&& {\rm STr}_s \big[ \sigma^s_3 Z (1 - \tilde{Z} Z)^{- 1} \sigma^s_3
  \tilde{Z} (1 - Z \tilde{Z})^{- 1} \big] \nonumber \\
&& \qquad = {\rm STr}_s \big[ \Sigma(Y, \tilde{Y})
  Q(\xi, \tilde{\xi}) \Lambda \Sigma'(Y, \tilde{Y}) Q(\xi, \tilde{\xi})
  \Lambda \big] \, .
\label{41}
\ea
The index $s$ indicates that the supertraces extend over superspace
only. The supermatrices $\Sigma$ and $\Sigma'$ are defined as
\ba
\Sigma(Y, \tilde{Y}) &=& \frac{1}{2}
   \left( \matrix{ (1 - Y \tilde{Y})^{- 1} & 0 \cr
               0 & (1 - \tilde{Y} Y)^{- 1} \cr} \right)
      \left( \matrix{ \sigma^s_3 & \sigma^s_3 Y \cr
     \tilde{Y} \sigma^s_3 & \tilde{Y} \sigma^s_3 Y \cr} \right) ,
      \nonumber \\
      \Sigma'(Y, \tilde{Y}) &=& \frac{1}{2}
               \left( \matrix{ (1 - Y \tilde{Y})^{- 1} & 0 \cr
               0 & (1 - \tilde{Y} Y)^{- 1} \cr} \right)
      \left( \matrix{ Y \sigma^s_3 \tilde{Y} & Y \sigma^s_3 \cr
        \sigma^s_3 \tilde{Y} & \sigma^s_3 \cr} \right) .
\label{40}
\ea
It is a very convenient feature of expression~(\ref{41}) that the
contributions from the universal mode $(Y, \tilde{Y})$ and from the
massive modes $(\xi, \tilde{\xi})$ factorize. That separation can be
carried a step further. We use the decomposition $Q(\xi_{b d},
\tilde{\xi}_{b d}) \Lambda = [Q(\xi_{b d}, \tilde{\xi}_{b d}) \Lambda
  - 1] + 1$. Supersymmetry must be broken for the same integration
variable in both the advanced and the retarded sector to obtain a
non-vanishing result. Therefore, terms linear in $( Q \Lambda - 1 )$
vanish upon integration, and the contribution of the massive modes to
the source terms is
\ba
&& \frac{\pi^2}{B^2} \bigg\{ \sum_{b d} {\rm STr}_s
\Big( \Sigma(Y, \tilde{Y}) [ Q(\xi_{b d}, \tilde{\xi}_{b d}) \Lambda
    - 1 ] \Sigma'(Y, \tilde{Y}) [ Q(\xi_{b d}, \tilde{\xi}_{b d})
    \Lambda - 1 ] \Big) \nonumber \\
&& \qquad + \sum_{b d} {\rm STr}_s \Big( \Sigma(Y, \tilde{Y})
  [ Q(\xi_{b d}, \tilde{\xi}_{b d}) \Lambda - 1 ] \Big) \nonumber \\
&& \qquad \qquad \times \sum_{b' d'} {\rm STr}_s \Big( \Sigma'(Y,
  \tilde{Y}) [ Q(\xi_{b' d'}, \tilde{\xi}_{b' d'}) \Lambda - 1 ] \Big)
\bigg\} .
\label{43}
\ea
It is our task to show that the integrals over the terms~(\ref{43})
with weight factor $\exp \{ - {\cal A}_{\rm bare}(\xi, \tilde{\xi})
\}$ and carried out for all $(\xi, \tilde{\xi})$ vanish for $B \to
\infty$.

\section{Evaluation}
\label{eva}

To argue in that direction we proceed as follows. We simplify the
source terms~(\ref{43}) by a suitable variable transformation. We
expand the bare effective action in powers of the new integration
variables. Terms up to second order define Gaussian
superintegrals. The exponential containing terms of higher order is
expanded in a Taylor series. We perform the Gaussian
superintegrals. We give an approximate estimate of the dependence on
$B$ of the terms so generated and on that basis argue that they vanish
for $B \to \infty$.

We display the procedure for a single contribution to the source
terms.  The procedure applies likewise to the remaining terms without
any additional difficulties and is not given here. We consider the
last term in expression~(\ref{43}). We introduce block notation,
writing
\be
\Sigma = \left( \matrix{ \Sigma_{+ +} & \Sigma_{+ -} \cr
                         \Sigma_{- +} & \Sigma_{- -} \cr} \right)
\label{43a}
\ee
and correspondingly for $\Sigma'$ and for $Q(\xi_{b d}, \tilde{\xi}_{b
  d})$.  It is convenient to write $\mu$ for $(b d)$ and $\nu$ for
$(b' d')$. For the pair $\Sigma_{+ +}$ $\Sigma'_{- -}$ the relevant
contribution is
\ba
&& \frac{\pi^2}{B^2} \bigg\langle \sum_\mu {\rm STr}_s \Big( \Sigma_{+ +}
   [ Q(\xi_\mu, \tilde{\xi}_\mu) \Lambda - 1 ]_{+ +} \Big) \nonumber \\
   && \qquad \times \sum_\nu {\rm STr}_s \Big( \Sigma'_{- -} [Q(\xi_\nu,
     \tilde{\xi}_\nu) \Lambda - 1 ]_{- -} \Big) \bigg\rangle \ . 
\label{44}
\ea
The angular brackets denote the superintegration over all $(\xi,
\tilde{\xi})$ with weight factor $\exp \{ - {\cal A}_{\rm bare} \}$.
According to Eqs.~(\ref{16}) we have $[Q(\xi, \tilde{\xi}) \Lambda -
  1]_{+ +} = 2 \xi \tilde{\xi} (1 - \xi \tilde{\xi})^{-1}$ and
$[Q(\xi, \tilde{\xi}) \Lambda - 1]_{- -} = 2\tilde{\xi} \xi (1 -
\tilde{\xi} \xi)^{- 1}$.

\subsection{Variable Transformation}

We simplify the form of the source term in expression~(\ref{44}) by
defining for each set of directed bond indices the variable
transformation
\be
\xi = \psi (1 + \tilde{\psi} \psi)^{- 1/2} , \quad
\tilde{\xi} = \tilde{\psi} (1 + \psi \tilde{\psi})^{- 1/2} ,
\label{45}
\ee
with inverse transformation
\be
\psi = \xi (1 - \tilde{\xi} \xi)^{- 1/2} , \quad
\tilde{\psi} = \tilde{\xi} (1 - \xi \tilde{\xi})^{- 1/2} .
\label{46}
\ee
Calculation shows that the Berezinian of the variable
transformation~(\ref{45}) is unity. Instead of the
constraints~(\ref{36}) we impose
\be
\sum_\mu \psi_\mu = 0 = \sum_\mu \tilde{\psi}_\mu \, .
\label{47}
\ee
To justify Eqs.~(\ref{47}) we observe that Eqs.~(\ref{36}) were
introduced in an ad-hoc fashion to guarantee that only $(2 B - 1)$ of
the variables $(\zeta_\mu, \tilde{\zeta}_\mu)$ and $(\xi_\mu,
\tilde{\xi}_\mu)$ are independent. Eqs.~(\ref{47}) serve that same
purpose.

From Eqs.~(\ref{45}) we have $[ Q(\xi, \tilde{\xi}) \Lambda - 1 ]_{+ +}
= 2 \psi \tilde{\psi}$ and $[ Q(\xi, \tilde{\xi}) \Lambda - 1 ]_{- -}
= 2 \tilde{\psi} \psi$. Expression~(\ref{44}) becomes
\ba
\frac{4 \pi^2}{B^2} \sum_{\mu \nu} \Big\langle {\rm STr}_s \big(
\Sigma_{+ +}\, \psi_\mu \tilde{\psi}_\mu \big) {\rm STr}_s \big(
\Sigma'_{- -}\, \tilde{\psi}_\nu \psi_\nu \big) \Big\rangle \ .
\label{48}
\ea
For the bare effective action of Eq.~(\ref{24}), the variable
transformation~(\ref{45}) leads to
\ba
{\cal A}_{\rm bare}(\psi, \tilde{\psi}) &=& + {\rm STr} \ln (1 + \psi
\tilde{\psi} ) \nonumber \\
&& + {\rm STr} \ln \big(1 - {\cal B} \psi (1 + \tilde{\psi}
\psi)^{- 1/2} {\cal B}^\dag \tilde{\psi} (1 + \psi \tilde{\psi})^{- 1/2}
\big) \nonumber \\
&=& {\rm STr} \big( \psi \tilde{\psi} - {\cal B} \psi {\cal B}^\dag
\tilde{\psi} \big) + \ldots 
\label{50}
\ea
where the dots indicate terms of higher order in $\psi$ and
$\tilde{\psi}$.

\subsection{Gaussian Superintegrals}
\label{gau}

In the expansion of ${\cal A}_{\rm bare}$ in Eq.~(\ref{50}), we retain
in the exponent only terms up to second order in $(\psi,
\tilde{\psi})$ (last line of Eq.~(\ref{50})). With ${\cal F}$ defined
in Eq.~(\ref{2}) these can be written as
\ba
{\cal A}_0 = \sum_{\mu \nu} {\rm Str}_s \big[ \psi_\mu (\delta_{\mu \nu} -
  {\cal F}_{\mu \nu}) \tilde{\psi}_\nu \big] .
\label{52}
\ea
Eq.~(\ref{52}) defines the Gaussian part ${\cal A}_0$ of the bare
effective action. The Perron-Frobenius operator ${\cal F}$ is
expanded in terms of its complex eigenvalues $\lambda_k$ and left and
right eigenvectors $\langle w_k |$ and $| u_k \rangle$ as
\be
{\cal F} = \sum_k | u_k \rangle \lambda_k \langle w_k | \, .
\label{53}
\ee
These (non-real) eigenvectors satisfy the relations $\langle w_k | u_l
\rangle = \delta_{kl}$. The matrix ${\cal F}$ is bistochastic, its
elements are positive or zero. The graph is connected. It follows from
the Perron-Frobenius theorem that there exists a non-degenerate
eigenvalue $\lambda_1 = + 1$. The associated left and right
eigenvectors $\langle w_1 |$ and $| u_1 \rangle$ have the components
$(1 / \sqrt{2 B}) \{ 1, 1, \ldots, 1 \}$. All other eigenvalues
$\lambda_i$ with $i \geq 2$ lie within or on the unit circle in the
complex plane. As stated below Eq.~(\ref{2}), we assume that all other
eigenvalues $\lambda_i$ with $2 \leq i \leq 2 B$ obey $|\lambda_i|
\leq (1 - a)$ with $a > 0$ even in the limit $B \to \infty$. The
matrix $| u_1 \rangle \langle w_1 |$ is an orthogonal
projector. Eqs.~(\ref{47}) guarantee that $| u_1 \rangle \langle w_1
|$ does not contribute to the sum on the right-hand side of
expression~(\ref{52}), confirming that the zero mode has been
eliminated. We emphasize that fact by defining the complementary
projector $ {\cal P} = 1 - | u_1 \rangle \langle w_1 |$, and by
writing expression~(\ref{52}) as
\ba
   {\cal A}_0 = \sum_{\mu \nu} {\rm STr}_s \Big[ \psi_{\mu} \Big(
     {\cal P} ( 1 - {\cal F} ) {\cal P}\Big)_{\mu \nu}
     \tilde{\psi}_{\nu} \Big] \, .
\label{54}
\ea
The bilinear form ${\cal A}_0$ defines the propagator of the
theory. The factor $\exp \{ - {\cal A}_0 \}$ defines Gaussian
superintegrals. The fundamental integral is
\be
\int {\rm d} (\psi, \tilde{\psi}) \psi_{\mu; s t} \tilde{\psi}_{\nu; t' s'}
\exp \{ - {\cal A}_0 \} = \delta_{s s'} \delta_{t t'} (-)^t \langle \mu |
{\cal P} (1 - {\cal F)}^{- 1} {\cal P} | \nu \rangle \ .
\label{55}
\ee
We have written $\mu = (b d)$ as before. The range of the superindices
$(s, t)$ is $(0, 1)$ or, equivalently, $(B, F)$.

The Taylor expansion of the exponential containing the dotted terms in
expression~(\ref{50}) generates products of supertraces each
containing powers of $\psi$ and $\tilde{\psi}$. For the Gaussian
integral over the product of these with the source term in
expression~(\ref{48}) we use the general result
\ba
&& \int {\rm d} (\psi, \tilde{\psi}) \prod_{i = 1}^n \psi_{\mu_i; s_i t_i}
\tilde{\psi}_{\nu_i; t_i' s_i'} \exp \{ - {\cal A}_0 \} \nonumber \\
&& \qquad = \prod_{i = 1}^n \sum_{\rm perm} \prod_{j = 1}^n \delta_{s_i s_j'}
\delta_{t_i t_j'} (-)^{t_i} \langle \mu_i | {\cal P} (1 - {\cal F})^{- 1}
      {\cal P} | \nu_j \rangle \ .
\label{56}
\ea
The sum is over all permutations of $(1, 2, \ldots, n)$. The
expressions generated by the Gaussian integrals~(\ref{56}) contain
products of factors $\langle \mu | {\cal P} (1 - {\cal F})^{- 1} {\cal
  P} | \nu \rangle$ and may become very lengthy. We use the
abbreviations
\be
W_{\mu \nu} = \langle \mu | {\cal P} (1 - {\cal F})^{- 1} {\cal P} |
\nu \rangle \ {\rm and, \ for} \ n > 1 \ , W^{(n)}_{\mu \nu} = \langle
\mu | {\cal P} (1 - {\cal F})^{- n} {\cal P} | \nu \rangle \ .
\label{62}
\ee

\subsection{Qualitative Estimation}
\label{est}

The integrals~(\ref{56}) generate products of matrix elements $W_{\mu
  \nu}$. Progress hinges on our ability to estimate the dependence of
these matrix elements and of sums of their products on the dimension
$(2 B)$ of directed-bond space for $B \to \infty$. Postponing a strict
treatment to future work, we here settle for the simple approximation
of using averages based upon the completeness relation.

For the diagonal elements we use Eq.~(\ref{53}), the relation $\langle
w_k | u_l \rangle = \delta_{kl}$, and the completeness relation and
find
\be
\langle \mu | {\cal P} (1 - {\cal F})^{- 1} {\cal P} | \mu \rangle
\approx \frac{1}{2 B} \sum_\mu \langle \mu | {\cal P} (1 - {\cal F})^{- 1}
        {\cal P} | \mu \rangle = \frac{1}{2 B} \sum_{k \geq 2} \frac{1}{1 -
          \lambda_k} \ .
\label{57}
\ee
Since ${\cal F}$ is real, the eigenvalues $\lambda_k$ are either real
or come in complex conjugate pairs. Therefore, the sum on the
right-hand side is real. By assumption, the eigenvalues obey
$|\lambda_k| \leq (1 - a)$ with $a > 0$. Therefore, the expression on
the right-hand side is positive and for all $B$ bounded from above by
$1 / a$. We accordingly estimate $W_{\mu \mu} = \langle \mu | {\cal P}
(1 - {\cal F})^{- 1} {\cal P} | \mu \rangle $ by $1 / a$. (Here and in
what follows we use the word ``estimate'' in the non-technical sense
of order-of-magnitude estimate). For terms with higher inverse powers
of $(1 - {\cal F})$ we find correspondingly $W^{(n)}_{\mu \mu} \approx
1 / a^n$. The factor $1 / a^n$ stems from the sum $\sum_k 1 / (1 -
\lambda_k)^n$. In the limit $B \to \infty$ that sum exists for all $n$
only if the gap in the spectrum of the Perron-Frobenius operator
${\cal F}$ does not close. That condition is also used in the
estimates given below.

For the non-diagonal elements we have
\be
\langle \mu | {\cal P} (1 - {\cal F})^{- 1} {\cal P} | \nu
\rangle (1 - \delta_{\mu \nu}) \approx \frac{1}{(2 B)^2}
\sum_{\mu \nu} \langle \mu | {\cal P} (1 - {\cal F})^{- 1} {\cal P} |
\nu \rangle (1 - \delta_{\mu \nu}) \ .
\label{58}
\ee
Since $\sum_\mu \langle \mu | {\cal P} = \sqrt{2 B} \langle w_1 |
{\cal P} = 0$ and $\sum_\mu {\cal P} | \mu \rangle = \sqrt{2 B} {\cal
  P} | u_1 \rangle = 0$, the first term of $(1 - \delta_{\mu \nu})$
gives a vanishing contribution. For the second term we use
Eq.~(\ref{57}) and find that the typical non-diagonal element $W_{\mu
  \nu}$ with $\mu \neq \nu$ is of order $1 / ((2 B) a)$. For
$W^{(n)}_{\mu \nu}$ with $\mu \neq \nu$ and $n > 1$ we correspondingly
have $W^{(n)}_{\mu \nu} \approx 1 / ((2 B) a^n)$.

After integration and use of the order-of-magnitude
estimate~(\ref{57}), the remaining terms in the expansion may carry a
product of matrix elements of the form $\prod_{i = 1}^n \langle
\mu_{i} | {\cal P} (1 - {\cal F})^{- 1} {\cal P} | \mu_{i + 1}
\rangle$. In addition to the steps taken in Eqs.~(\ref{57}) and
(\ref{58}) we use the approximation $| \mu_i \rangle \langle \mu_i |
\approx (1 / (2 B)) \sum_\mu | \mu \rangle \langle \mu |$ for the
intermediate projectors $| \mu_i \rangle \langle \mu_i |$ with $2 \leq
i \leq n$. That gives
\ba
\prod_{i = 1}^n \langle \mu_{i} | {\cal P} (1 - {\cal F})^{- 1}
     {\cal P} | \mu_{i + 1} \rangle = \prod_{i = 1}^n W_{\mu_i \mu_{i + 1}}
       \approx \frac{1}{a^n (2 B)^{n - 1}} \bigg( \delta_{\mu_1 \mu_{n + 1}}
       - \frac{1}{2 B} \bigg) \ .
\label{61}
\ea
The relation~(\ref{61}) holds likewise (with appropriate changes of
the power of $1 / a$) in cases where one or several of the
denominators $(1 - {\cal F})$ carry powers larger than unity.

For an estimate of the large-$B$ dependence of expressions involving
${\cal B}$ or ${\cal B}^\dag$, we use the unitarity relation
$\sum_{\sigma = 1}^{2 B} {\cal B}^\dag_{\rho \sigma} {\cal
  B}^{}_{\sigma \rho} = 1$. It implies that in the ergodic limit we
have
\be
| {\cal B}_{\mu \nu}| \approx \frac{1}{\sqrt{2 B}} \ .
\label{59}
\ee

The estimates~(\ref{56}), (\ref{61}) and (\ref{59}) provide us with
the tools needed to give an order-of-magnitude estimate of the
$B$-dependence of the source terms and the terms generated by the
Taylor expansion of the higher-order terms in ${\cal A}_{\rm bare}$.

\subsection{Source Term}

In expression~(\ref{48}) we first disregard contributions due to the
Taylor expansion of higher-order terms in ${\cal A}_{\rm
  bare}$. Eq.~(\ref{56}) implies that the Gaussian integrals over
products of factors $(\psi, \tilde{\psi})$ lead to pairwise
contractions. In the source term~(\ref{48}) that results in the sum of
two terms,
\ba
&& \bigg\langle \sum_{t_1} \psi_{\mu, {s_1} {t_1}} \tilde{\psi}_{\mu, {t_1}
  {s'_1}} \sum_{s_2} \tilde{\psi}_{\nu, {t_2} {s_2}} \psi_{\nu, {s_2} {t'_2}}
\bigg\rangle \nonumber \\
&& \qquad = \bigg\langle \sum_{t_1} \psi_{\mu, {s_1} {t_1}}
\tilde{\psi}_{\mu, {t_1} {s'_1}} \bigg\rangle \ \bigg\langle \sum_{s_2}
\tilde{\psi}_{\nu, {t_2} {s_2}} \psi_{\nu, {s_2} {t'_2}} \bigg\rangle
\nonumber \\
&& \qquad \qquad + (-)^{s_1 + t_2} \sum_{t_1} \sum_{s_2} \bigg\langle
\psi_{\mu, {s_1} {t_1}} \tilde{\psi}_{\nu, {t_2} {s_2}} \bigg\rangle \
\bigg\langle \tilde{\psi}_{\mu, {t_1} {s'_1}} \psi_{\nu, {s_2} {t'_2}}
\bigg\rangle \ .
\label{63}
\ea
According to Eq.~(\ref{56}) the first term vanishes, and the second
term equals $W_{\mu \nu} W_{\nu \mu} \delta_{s_1 s'_1} \delta_{t_2
  t'_2}$. We have $|\sum_{\nu} W_{\mu \nu} W_{\nu \mu}| \leq || W W
||_{\rm op} \leq || W ||^2_{\rm op} \leq a^{- 2}$. The double bars
denote the operator norm. Hence $|\sum_{\nu} W_{\mu \nu} W_{\nu \mu}|
\leq (2 B) / a^2$. The term~(\ref{48}) carries the factor $1 / B^2$.
For $B \to \infty$ it vanishes as $1 / B$.

\subsection{Higher-Order Contributions}
\label{oth}

Among the terms in expression~(\ref{48}) that arise from the Taylor
expansion of higher-order terms in ${\cal A}_{\rm bare}$ in
Eq.~(\ref{50}), we first address terms that originate from ${\rm STr}
\ln (1 + \psi \tilde{\psi})$. Except for numerical factors the general
term in the Taylor expansion has the form
\be
\frac{4 \pi^2}{B^2} \Big\langle \sum_{\mu \nu} {\rm STr}_s \big(
\Sigma_{+ +}\, \psi_\mu \tilde{\psi}_\mu \big) {\rm STr}_s \big(
\Sigma'_{- -}\, \tilde{\psi}_\nu \psi_\nu \big) \prod_{i = 1}^m {\rm
  STr} (\psi \tilde{\psi})^{n_i} \Big\rangle \ ,
\label{64}
\ee
with integer $m \geq 1, n_i \geq 2$. Gaussian integration as in
Eq.~(\ref{56}) leads to pairwise contraction of all factors $\psi$ and
$\tilde{\psi}$ and, thus, to $k = 2 + \sum_i n_i$ factors $W$. A
non-vanishing result is obtained only if supersymmetry is violated in
every supertrace in expression~(\ref{64}). Therefore, all supertraces
in expression~(\ref{64}) must be linked by pairwise
contractions. That, incidentally, is the reason why in Eq.~(\ref{63})
the first term on the right-hand side vanishes.

We consider two examples. For $m = 1, n_1 = 2$, the nonvanishing links
yield two terms,
\ba
&& (1 / B^2) \sum_{\mu \nu \rho} W_{\mu \nu} W_{\nu \rho}
W_{\rho \mu} W_{\rho \rho} = (1 / B^2) \sum_\rho W^{(3)}_{\rho \rho}
W_{\rho \rho} \ , \nonumber \\
&& (1 / B^2) \sum_{\mu \nu \rho} W_{\mu \rho} W_{\rho \mu} W_{\rho \nu}
W_{\nu \rho} = (1 / B^2) \sum_\rho W^{(2)}_{\rho \rho} W^{(2)}_{\rho \rho} \ .
\label{65}
\ea
In the first expression we use $W_{\rho \rho} \approx 1 / a$,
$W^{(3)}_{\rho \rho} \approx 1 / a^3$. The sum over $\rho$ yields a
factor $(2 B)$. Altogether the term is estimated as $1 / ((2 B) a^4)$
and vanishes for $B \to \infty$. In the second expression we use
$W^{(2)}_{\rho \rho} \approx 1 / a^2$ and find the same result. For $m
= 2, n_1 = 2 = n_2$ we correspondingly obtain the terms
\ba
&& (1 / B^2) \sum_{\rho \tau} W^{(3)}_{\rho \rho} W_{\rho \tau}
W_{\tau \rho} W_{\tau \tau} \ , \ (1 / B^2) \sum_{\rho \tau}
W^{(3)}_{\tau \rho} (W_{\rho \tau})^2 W_{\tau \rho} \ , \nonumber \\
&& (1 / B^2) \sum_{\rho \tau} W^{(3)}_{\tau \rho} W_{\rho \tau} W_{\rho \rho}
W_{\tau \tau} \ , \ (1 / B^2) \sum_{\rho \tau} W^{(2)}_{\rho \tau}
W^{(2)}_{\tau \rho} W_{\rho \tau} W_{\tau \rho} \ , \nonumber \\
&& (1 / B^2) \sum_{\rho \tau} W^{(2)}_{\rho \rho}
W^{(2)}_{\tau \tau} W_{\rho \tau} W_{\tau \rho} \ , \ (1 / B^2)
\sum_{\rho \tau} W^{(2)}_{\rho \tau} W^{(2)}_{\tau \rho} W_{\rho \rho}
W_{\tau \tau} \ .
\label{65a}
\ea
We use the estimates~(\ref{57}, \ref{61}). The terms of leading order
are the ones where two factors $W^{(n)}$ with $n = 1$ or $n = 2$ or $n
= 3$ each carry identical indices. These are the terms number 1, 3, 5,
6. They vanish asymptotically as $1 / (2 B)$.

From these examples we deduce the following rules. (i) In
expression~(\ref{64}), contraction of the four factors $(\psi_\mu,
\tilde{\psi}_\mu)$ and $(\psi_\nu, \tilde{\psi}_\nu)$ with one another
and with corresponding factors in the product over traces yields a
nonvanishing result only if it generates either two factors $W^{(2)}$
or a single factor $W^{(3)}$. These carry indices of the factors in
the product over traces. For every pair $(\rho, \sigma)$, the factors
$W^{(2)}_{\rho \sigma}$ and $W^{(3)}_{\rho \sigma}$ are, both for
$\rho = \sigma$ and for $\rho \neq \sigma$, of the same order in $1 /
(2 B)$ as the factor $W_{\rho \sigma}$. For purposes of counting
powers of $(2 B)$ it suffices, therefore, to disregard the four
factors $(\psi_\mu, \tilde{\psi}_\mu)$ and $(\psi_\nu,
\tilde{\psi}_\nu)$ together with the summation over $\mu$ and $\nu$
and to consider only contractions within the product of supertraces.
(ii) The terms of leading order in $1 / (2 B)$ are obtained by
contracting, in every supertrace ${\rm STr} (\psi \tilde{\psi})^{n_i}$,
$(n_i - 1)$ pairs $(\psi_\rho, \tilde{\psi}_\rho)$ with one another.
That is because contraction of pairs of $\psi$'s in the same (in
different) supertraces generates factors $W_{\rho \rho} \propto 1 / a$
($W_{\rho \sigma} \propto 1 / (2 B a)$ with $\rho \neq \sigma$,
respectively). Contraction of pairs within the same supertrace
generates $(n_i - 1)$ factors $W_{\rho \rho}$ and reduces each
supertrace to $(1 / a^{n_i - 1} ) {\rm STr} (\psi \tilde{\psi})$.
These rules imply that the only remaining contractions in
expression~(\ref{64}) are over the term
\be
[{\rm STr} (\psi \tilde{\psi})]^m \to \sum_{\mu_1, \ldots, \mu_m}
W_{\mu_1 \mu_2} \times \ldots \times W_{\mu_m \mu_1} = \sum_{\mu_1}
W^{(m)}_{\mu_1} \ .
\label{65b}
\ee
To the extent that our approximations (of replacing fluctuating
quantities by their averages) capture the qualitative aspects of the
problem, we may conclude that expression~(\ref{64}) is of order $1 /
(2 B)$ and vanishes for $B \to \infty$.

The general term in the Taylor expansion contains, in addition to the
terms in expression~(\ref{64}), also products of supertraces that
contain the elements of ${\cal B}$. Each such supertrace has the form
${\rm STr} [{\cal B} \psi (\tilde{\psi} \psi)^m {\cal B}^\dag
  \tilde{\psi} (\psi \tilde{\psi})^n]^l$ with $l \geq 1$ and $(m, n)$
zero or positive integer, see Eq.~(\ref{50}). In contracting factors
$\psi$ and $\tilde{\psi}$ we apply rule (ii). The terms of leading
order are obtained by replacing in each supertrace $(\tilde{\psi}
\psi)^m$ by $1 / a^m$ and $(\psi \tilde{\psi})^n$ by $1 / a^n$. We
proceed likewise for the product of supertraces in
expression~(\ref{64}). That reduces the general expression to
\ba
&& \frac{1}{B^2} \Bigg\langle \sum_{\mu \nu} {\rm STr}_s \big(
\Sigma_{+ +}\, \psi_\mu \tilde{\psi}_\mu \big) {\rm STr}_s \big(
\Sigma'_{- -}\, \tilde{\psi}_\nu \psi_\nu \big) \nonumber \\
&& \ \ \  \times \Bigg\{ [{\rm STr}
  (\psi \tilde{\psi})]^m \prod_{i = 1}^k {\rm STr} \bigg( {\cal B}
\psi {\cal B}^\dag \tilde{\psi} \bigg)^{l_i} \Bigg\} \Bigg\rangle \ .
\label{66}
\ea
Here $m \geq 0$ and $l_i \geq 1$ for $i = 1, \ldots, m$. We require $k
\geq 1$ as the case $k = 0$ has been considered above. We apply rule
(i) and disregard the difference between factors $W^{(n)}$ and $W$ in
estimating the dependence of expression~(\ref{66}) on $1 / (2 B)$. In
other words, we confine ourselves to contractions involving elements
$(\psi, \tilde{\psi})$ in the curly brackets in expression~(\ref{66}).
For $m = 0$, all supertraces carrying elements of ${\cal B}$ must be
connected by pairwise contractions. That generates $\sum_i l_i$
factors $W$. Each such factor carries a pair of summation indices. No
two summation indices in any of these factors are the same. Therefore,
each factor $W$ is of order $1 / (2 B)$. There are $\sum_i l_i$
elements of ${\cal B}$ and of ${\cal B}^\dag$, each of order $1 /
\sqrt{2 B}$.  There are $2 \sum_i l_i$ independent summations over
directed-bond space. Together with the prefactor $1 / B^2$, we
therefore expect expression~(\ref{66}) to be of order $1
/ (2 B)^2$ for $m = 0$.

That expectation remains unaltered for $m > 0$. We show that first for
$m = 1$, writing ${\rm STr} (\psi \tilde{\psi}) = \sum_\rho {\rm
  STr}_s (\psi_\rho \tilde{\psi}_\rho)$. Contraction links the factor
$\psi_\rho$ (the factor $\tilde{\psi}_\rho$) with some factor
$\tilde{\psi}_\sigma$ (with some factor $\psi_\tau$,
respectively). Both $\tilde{\psi}_\sigma$ and $\psi_\tau$ occur in the
product over $i$ in expression~(\ref{66}). The result is $\sum_\rho
W_{\rho \sigma} W_{\tau \rho} = W^{(2)}_{\tau \sigma}$ while for $m =
0$ contracting $\tilde{\psi}_\sigma$ and $\psi_\tau$ gives $W_{\tau
  \sigma}$. Since $W_{\tau \sigma}$ and $W^{(2)}_{\tau \sigma}$ are of
the same order, the expressions~(\ref{66}) with $m = 0$ and with $m =
1$ are of the same order, too. The argument can straightforwardly be
extended to $m \geq 2$. This concludes our heuristic reasoning that
the general term containing matrix elements of ${\cal B}$ vanishes for
$B \to \infty$ as $1 / (2 B)^2$.

\section{Summary and Discussion}

We have given a brief account of an approach to chaotic quantum graphs
that aims at demonstrating the BGS conjecture using supersymmetry and
the color-flavor transformation. We have paid particular attention to
the treatment of the massive modes as these are defined in a coset
space. We have used the assumption that the spectrum of the
Perron-Frobenius operator possesses a finite gap even for infinite
graph size. We have shown that the effective action and the source
terms are given by Eqs.~(\ref{24}) and (\ref{38}) and by
Eq.~(\ref{43}), respectively.

To evaluate the resulting generating function, we have defined
Gaussian superintegrals by expanding the effective action up to terms
of second order. The remaining terms are expanded in a Taylor
series. We have carried out the Gaussian superintegrals over the
product of that series with the source terms. We have given
order-of-magnitude estimates of the resulting expressions using
averages based upon completeness and unitarity. Assuming that these
approximations are valid, we are led to the conclusion that the
contribution of massive modes to the two-point function vanishes for
large graph size. Therefore, that function attains universal form.

The rough estimates in Section~\ref{est} are based on averages and
require small fluctuations. That may be unsatisfactory. We are working
on strict estimates. We hope to be able soon to report on these,
combining that with a more detailed account of conceptual and
technical aspects of the steps taken in Section~\ref{eff} that were
treated here only cursorily.

The author is much indebted to M. R. Zirnbauer. Without his numerous
useful suggestions, especially concerning the developments in
Section~\ref{eff}, this contribution would not have come into
existence.

\end{document}